\documentstyle[11pt,newpasp,twoside,epsf]{article}
\def\edcomment#1{\iffalse\marginpar{\raggedright\sl#1\/}\else\relax\fi}
\reversemarginpar

\begin{document}
\title{Star Formation in Turbulent Molecular Clouds}
 \author{Andreas Burkert}
\affil{Max-Planck-Institut for Astronomy, K\"onigstuhl 17, D-69117 Heidelberg, Germany}

\begin{abstract}
Recent progress in the understanding of star formation is summarized. A consistent picture is emerging
where molecular clouds form with turbulent velocity fields and clumpy substructure,
imprinted already during their formation. The clouds are initially supported by supersonic
turbulence which dissipates however within massive clumps on short timescales,
of order their local dynamical timescales.  As a result, the clumps collapse and fragment into 
stellar clusters.
Subsequent energetic feedback processes by the newly formed stars
will change the thermal and kinematical state of the surrounding molecular gas, triggering or
suppressing star formation and leading eventually to the disruption of the molecular clouds.
More work is required to understand these processes in greater details.
\end{abstract}

\section{Introduction}

Star formation is one of the most important and yet unsolved problems 
of astrophysics. 
It plays an important role in many different areas of astrophysics,
especially in the field of galaxy formation and evolution.
For example, recently Hodgkin et al. (2000) and Ibata et al. (2000) found very old white
dwarfs with kinematical properties similar
to that of galactic halo stars, which might have formed from a progenitor halo population with
several times the mass of the galactic disk. Without a theory of star formation
it will be difficult to understand the origin of this component 
which indicates a yet unknown phase of violent and efficient star formation with a
peculiar IMF during the
epoch of protogalactic collapse.

A star forms as the final stage of various complex, non-linear dissipative processes that
are the result of the interplay between turbulent, thermal and magnetic pressure effects
on one side and gravity on the other side. Very different scales are envolved with
densities that vary by more than 10 orders of magnitudes and with complex, probably fractal
substructures (Efremov \& Elmegreen 1998).
Once a star has formed it can trigger and suppress further star formation.  On galactic
scales feedback by local star formation can couple in a complex, non-linear way with galactic dynamics,
leading in some cases to a self-regulated and
low global star formation rate or, in starburst situations, to a runaway star formation process.

Although a consistent star formation theory is still missing, recent high-resolution observations
and numerical simulations have provided new and interesting insight into several important aspects 
of star formation. This paper will summarize some of these
new results.  An excellent and detailed description of our current knowledge of star formation is provided by
the proceedings of Protostars and Planets IV (2000).

\section{Molecular clouds as sites of star formation}

All stars are known to form in molecular clouds.
The global distribution of molecular gas in the Milky Way
is determined using tracer molecules like CO or mm continuum dust emission (Scoville \& Solomon 1975). 
More dense cloud regions are observed in high-excitation lines or transitions of molecules with large dipole moments
(Benson \& Myers 1989).  These observations show that
molecular clouds are one of the most massive objects
in galaxies with masses $M \approx 10^4 - 10^6 M_{\odot}$, radii $R \approx 5 - 30$ pc,
densities $n \approx$ 100 cm$^{-3}$ and temperatures of order 10 -- 30 K, corresponding to internal
isothermal sound speeds of $c_s \approx$ 0.3 km/s (Blitz 1993; Williams, Blitz \& McKee 2000). 
There exists a sharp upper cutoff in the cloud mass
distribution at $6 \times 10^6 M_{\odot}$ (Williams \& McKee 1997) which is not explained up to now.
Spectroscopic measurements indicate a
turbulent kinematic state with highly supersonic, irregular gas velocities of $\sigma \approx$ 3 km/s which
very likely have a strong influence on cloud stability and dynamics  (Larson 1981; Heithausen 1996; Myers \& Gammie 1999). 
Zeeman measurements give typical magnetic field strengths that are of the order of a few tens of microgauss
(Crutcher 1999).

Given the fact that clouds are strongly Jeans unstable, most of them should be
in a state of global gravitational contraction or collapse which is not observed. 
This leads to the conclusion that turbulent 
and magnetic pressure can compensate the gravitational force, leading to a global 
virial equilibrium state.  Observations (Crutcher 1999) show that the ratio of turbulent energy 
$E_{turb}$ to potential energy $E_{pot}$ in molecular clouds is

\begin{equation}
\frac{E_{turb}}{E_{pot}} \approx 2.5 \left(\frac{R \sigma^2}{GM} \right) \approx 0.25
\end{equation}

\noindent which is close to the virial value of 0.5. In addition there exists
energy equipartition between the turbulent energy and the magnetic energy $E_B$ 
in clouds (Crutcher 1999),

\begin{equation}
\frac{E_B}{E_{turb}} = \frac{2}{9} \frac{B^2R^3}{M \sigma^2} \approx 0.5
\end{equation}

\noindent indicating that magnetic fields are not negligible in regulating cloud dynamics.
They are however also not strong enough to magnetically stabilize molecular clouds 
against gravitational collapse.

That molecular clouds are the sites of star formation is not very 
surprising as
gravitationally unstable regions with masses of order a stellar mass
require a very cold and dense environment. It is however interesting that
half of the mass of the interstellar medium within the solar circle,
that is $M_{H_2} \approx 1 - 2 \times 10^9 M_{\odot}$ is in  molecular
clouds (Dame et al. 1993). Given such a large reservoir of star forming material,
the galactic star formation rate SFR is remarkably low 
(Scalo 1986, Evans 1999, Pringle et al. 2000):

\begin{equation}
SFR = \eta_{SF} \left( \frac{M_{H_2}}{\tau_{SF}} \right) \approx 2 - 5 M_{\odot}/yr
\end{equation}

\noindent where $\eta_{SF}$ is the star formation efficiency, that is the total mass of stars that form
per total molecular cloud mass and $\tau_{SF}$ is the star formation timescale. Adopting a reasonable star
formation efficiency $\eta_{SF} \approx 0.1$ would require a very long star formation timescale of order
$\tau_{SF} \approx 10^8$ yrs which is 10 -- 50 times 
the internal dynamical timescale of dense molecular clouds.
Another possibility is a very small star formation efficiency $\eta_{SF} \leq 0.01$ or a combination 
of both effects. Note, that a low star formation efficiency does not necessarily require that the molecular material 
is ionized and
heated. It could also be dispersed by supernovae, 
remaining in a cold, molecular state but with a small density that is not
large enough for stars to form (Pringle et al. 2000).

The formation of molecular clouds is not well understood up to now. Large-scale density waves
in spiral galaxies might lead to cooling of the warm, low-density intercloud medium (n $\approx$ 0.1 cm$^{-3}$).
Subsequent thermal instabilities in the compressed atomic gas could
create clumpy substructures on scales of order 0.1 pc with turbulent
supersonic velocities (Burkert \& Lin 2000,V\'{a}zquez-Semadeni et al. 2000). In order to form a $10^6 M_{\odot}$ cloud out of atomic hydrogen with
initial density of 1 cm$^{-3}$, gas must be accumulated from a cube of 0.4 kpc box size.
After this region has broken up into many, initially Jeans stable 
clumps their subsequent coagulation and contraction could lead to a dense cloud.
This process would occur  on timescales
of order the dynamical timescale of the interclump medium which corresponds to several $10^7$ yrs.
The newly formed molecular cloud would initially be dominated and stabilized by the 
turbulent clump motions that act like an additional kinetic pressure term. However, during the late contraction phase
the turbulent energy should dissipate, leading to gravitational collapse and star formation. A 
similar scenario has been suggested by Pringle et al (2000) who however considered the hypothesis 
that molecular clouds form by agglomeration of an already cold but invisible
molecular phase of the interstellar medium.

\section{The density and velocity structure of molecular clouds}

Molecular clouds are highly irregular, clumpy and filamentary (Mizuno et al. 1995). 
Substructure, i.e. clumps can be identified from spectral line maps of
molecular emission as coherent regions in the l-b-v-space. The clumps follow a power-law mass
distribution N(m) $\sim m^{-0.5}$ with no characteristic scale (Elmegreen 1997), consistent
with a fractal distribution. Cloud boundaries also show a complex structure with
a projected fractal dimension D=1.4 (Scalo 1990; Falgarone et al. 1991) as determined from their
perimeter-area relation. 
As the average clump density is $n_{clump} \approx 10^3 cm^{-3}$ which is
10 times the mean gas density in clouds, the clump volume filling factor must be small, 
of order 10\% (Williams et al. 1994). Blitz (1990, 1993) argues that atomic gas pervades
the interclump medium. In general, 
the nature and thermal state of the interclump medium which fills 90\% of the volume in clouds and which
might play an important role in confining clumps is still unclear.

In 1981 Larson found interesting correlations between the clump densities, their internal velocity dispersions
and their sizes: $\rho_c \sim R_c^{-1.1}$ and $\sigma_c \sim R_c^{0.4}$. These Larson relations have been
interpreted as a signature that molecular clumps are stable and in virial equilibrium with a balance  between
gravity and internal turbulent pressure support which would require $\rho_c R_c^2 \sim \sigma_c^2$, 
in agreement with the observations. However, recent analytical calculations (Myers \& Gammie 1999)
and numerical simulations ((V\'{a}zquez-Semadeni et al. 1997; Stone 1999) 
of turbulent molecular clouds indicate that the observed linewidth-size-relationship resembles a
Kolmogorov spectrum which is characterized by a spectral density of turbulent energy E(k) that depends on wave number k
as E(k) $\sim k^{-5/3}$ for incompressible hydrodynamical turbulence (Kolmogorov 1941).
The interstellar medium his however a highly compressible 
gas. In this case, Kornreich and Scalo (2000) note that the spectrum of compressible modes should
approach a Burger's (1974) spectrum of the form E(k) $\sim k^{-2}$, which results simply from the Fourier structure
of a system of shocks. 
Although this interpretation is promising it fails to reproduce the observed density-size relationship of clumps.

\begin{figure}
\plotfiddle{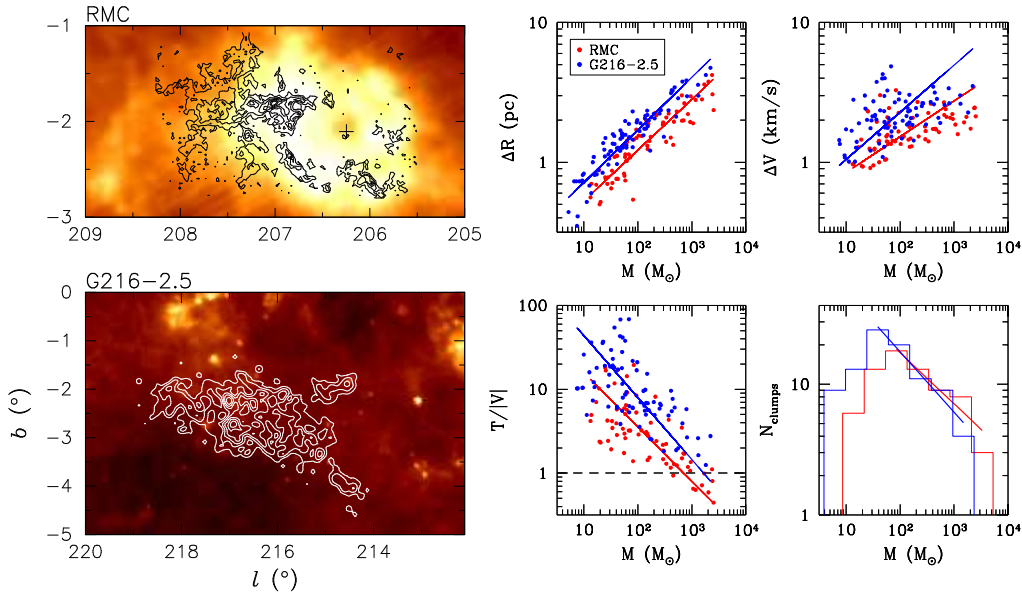}{8.0 cm}{90}{50}{50}{200}{-50}
\caption{Structure in the Rosette and G216-2.5 clouds. The left panels
show contours of velocity-integrated CO emission, overlaid on a grayscale image
of the IRAS 100-$\mu$m intensity. The four rightmost panels show
power-law relations between clump mass and size, linewidth, the ratio
between kinetic energy T and potential energy W and the clump mass spectrum.
Figure from Williams et al. (2000).}
\end{figure}

That clumps are actually transient and unbound density fluctuations instead of bound and self-gravitating 
has been demonstrated by Williams et al. (1994)
who searched for differences in clump structure between a star forming cloud (Rosette molecular cloud)
and a non-starforming cloud (G216-2.5, see Maddalena \& Thaddeus 1985). They detected of order 100 clumps
in each cloud which follow power-law relations  in their global properties, close to the Larson 
laws. However all clumps
have ratios of internal turbulent energy to gravitational energy that are much larger than unity.
G216-2.5 completely lacks bound clumps. 
In the Rosette molecular cloud there do however exist a few very
massive clumps with masses of order $10^3 M_{\odot}$ that  seem to be gravitationally bound and that
might be able to  collapse and  form a small stellar cluster. These interesting observations  
indicate that
turbulence does not necessarily arise by internal star formation as otherwise one would expect to
find less turbulent and more gravitationally bound clumps in the non-star forming region, in contradiction
with the observations.

\section{Driven molecular cloud turbulence}

As discussed in section 2, 
the relatively low star formation rate in the Milky Way (equation 3) indicates molecular cloud lifetimes
that are one order of magnitude longer (Blitz \& Shu 1980) than their dynamical timescale. 
In this case, the clouds must be stabilized
against collapse by their turbulent or magnetic pressures. Recently, a large number of numerical
models (Mac Low et al. 1998, Mac Low 1999, Stone et al. 1998) have clearly shown that 
turbulence can stabilize clouds, however only for a dynamical timescale due to efficient dissipation
of kinetic energy even in the presence of strong magnetic fields.
Turbulence must therefore be constantly driven to keep a cloud in a stable highly turbulent state for
10 dynamical timescales. A natural energy source would be newly formed stars. This question has been investigated
by Klessen et al. (2000) who demonstrated that the stability of turbulent
clouds depends not only on the global energy input rate but also on the wavelength  of
driving. Local collapse and fragmentation can only be halted if the driving scale is smaller than the local
Jeans length which is of order 0.1 pc. Roughly $10^6$ energy sources would be required in order to maintain
the supersonic velocity field for more than a dynamical timescale  and prevent global collapse and cluster formation.
Protostars can be
ruled out as local drivers of turbulence as one would need a massive 
stellar cluster to prevent the cloud from
collapsing and condensing into such a cluster. Whether there exist other turbulent drivers in molecular clouds
is still not known.

One  solution to the driving problem is the possibility that clouds are not stabilized by turbulence
but rather are transient structures that form on long timescales but evolve quickly (Elmegreen 2000)  and condense into
stars on their local dynamical timescale. Efremov \& Elmegreen (1998) find that supersonic turbulence and
the Larson relations are not restricted to the cold molecular gas component but are general properties
of the interstellar medium in galaxies. In this case, the substructure of molecular clouds could have
been imprinted not by internal turbulent driving but already during their formation as discussed in section 2.

\section{From Clouds to Stellar Clusters}

Stars form through the gravitational collapse of dense molecular clouds cores which can be identified
in optically thin lines. The cores have masses of order 5 $M_{\odot}$ and radii of order
0.1 pc (Myers \& Benson 1983). 
In contrast to clumps, they are gravitationally bound structures that are supported by thermal pressure
with subsonic internal turbulent motions (Barranco \& Goodman 1998). 
There exists a clear break in self-similarity on the scale of molecular cloud cores
which is similar to the 
local Jeans length and where thermal pressure effects begin to dominate. 
Cores do not follow the scale-free relations of clumps but instead
have mass functions that deviate strongly from a power-law distribution and that are very similar
to the stellar initial mass function (Motte et al. 1998; Testi \& Sargent 1998). We can
conclude that cores represent the direct progenitors of stars. Understanding the formation
of cloud cores and their mass spectrum
might also explain the origin of the stellar initial mass function.

\begin{figure}
\plotone{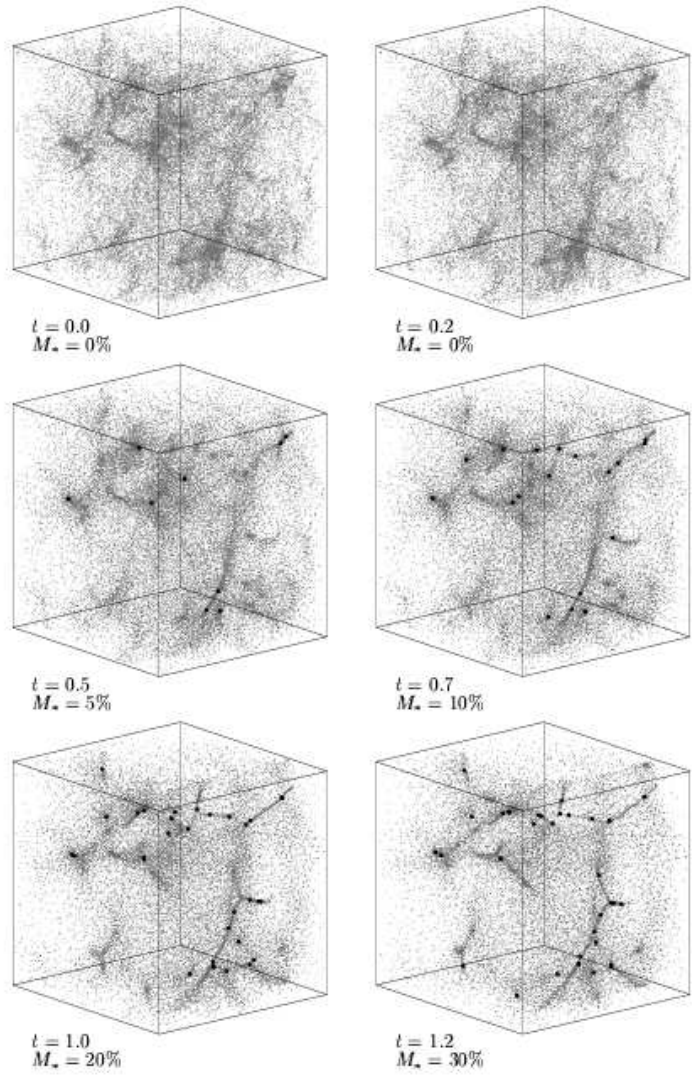}
\caption{Snapshots of the condensation of an initially 
turbulent gas cloud into a stellar cluster.
During the evolution (t in units of the local
dynamical time) Jeans-unstable clumps start to collapse and form compact
cores which are represented by sink particles in this SPH simulation. 
M$_*$ denotes the percentage of the total gas mass accreted onto
protostellar cores.}
\end{figure}

Observations indicate that stellar clusters form on short timescales of order $10^6$ yrs. Such an event can be
triggered within a massive clump if it becomes self-gravitating  and dissipates its internal
turbulent energy on the local dynamical timescale. Numerical simulations by Klessen \& Burkert (1998,2000) have
investigated this process in detail (Fig. 2). 
Their models lead to a cluster of cores with log-normal mass functions, resembling
the stellar initial mass function.
The models also show that cores that form in turbulent clumps are gravitationally unstable
regions where gravity has overcome the internal turbulent pressure. These cores collapse while still accreting
material from the surrounding. This seems to be in contradiction with recent observations by 
Alves et al (2001) who investigated the internal density distribution of cores and found that they
follow a stable Bonnor-Ebert distribution with no signature of gravitational collapse.
The origin of stable, dense cores in turbulent clouds is not known up to now.

The final stage of star formation is the collapse of a molecular core and the formation of
a quasi-static stellar core. Most stars form as binaries or multiple systems with a very broad period distribution, ranging
from one day to $10^{10}$~days and eccentricities, ranging from~0 to~1. 
The origin of this distribution is not well understood. Certainly, the 
specific angular momentum distribution of cloud cores must play an important role. 
Rotational properties of cores with subsonic turbulent velocity fields
have been investigated by Burkert \& Bodenheimer (2000) who showed  that random 
superpositions of the nonthermal, turbulent velocity modes with a Kolmogorov spectrum 
can explain the observed line-of-sight velocity gradients
and inferred spin distributions of cores (Barranco \& Goodman 1998). 
Numerous simulations have investigated the collapse and fragmentation
of disturbed, rotating cores (for a summary see Bodenheimer et al. 2000). 
Multiple systems seem to form naturally in these simulations, however the details
are unclear due to numerical resolution problems.

\section{Summary}
Our insight into the process of star formation has increased substantially over the past
years. Many questions are  however still unsolved:

\begin{itemize}
\item How do turbulent, clumpy molecular clouds form?
\item What is the timescale for molecular cloud formation?
\item On which timescale do clouds condense into stellar clusters?
\item What is the nature of the interclump medium in molecular clouds?
\item Is turbulence in molecular clouds driven or decaying?
\item If turbulence is driven, what is its driver?
\item How do molecular cloud cores from inside molecular clouds?
\item Are cloud cores initially gravitationally stable?
\item How does the broad period distribution of binary systems arise?
\item How important are magnetic fields in regulating the star formation process?
\end{itemize}

Given the fast progress in the field of star formation both on the observational and theoretical side,
it seems likely that we will be able to answer many of these questions within the next years.

\end{document}